\documentclass{XrU2005}
\usepackage{epsfig}

\title{Results from the First {\it INTEGRAL} AGN Catalogue}
\author{Volker Beckmann$^{1,2}$, Simona Soldi$^{3,4}$, Chris R. Shrader$^{1}$ and Neil Gehrels$^{1}$}
\affil{$^1$NASA Goddard Space Flight Center, Exploration of the Universe Division, Code 661, Greenbelt, MD 20771, USA}
\affil{$^2$Joint Center for Astrophysics, Department of Physics, University of Maryland, Baltimore County, MD 21250, USA}
\affil{$^3$INTEGRAL Science Data Centre, 16 chemin d'\'Ecogia, 1290 Versoix, Switzerland}
\affil{$^4$Observatoire de Gen\`eve, 51 chemin des Maillettes, 1290 Sauverny, Switzerland}

\newcommand{\lae}{\mathrel{<\kern-1.0em\lower0.9ex\hbox{$\sim$}}}

\begin{document}

\keywords{galaxies: active, catalogues, gamma rays: observations, X-rays: galaxies, galaxies: Seyfert}

\maketitle

\begin{abstract}
We present results based on the first {\it INTEGRAL} AGN catalogue. The catalogue includes 42 AGN, of which 10 are Seyfert~1, 17 are Seyfert~2, and 9 are intermediate Seyfert~1.5. The fraction of blazars is rather small with 5 detected objects, and only one galaxy cluster and no star-burst galaxies have been detected so far. 
The sample consists of  bright ($f_X > 5 \times 10^{-12} \rm \, erg \, cm^{-2} \, s^{-1}$), low luminosity (${\bar L_X} = 2 \times 10^{43} \rm \, erg \, s^{-1}$), local (${\bar z} = 0.020$) AGN.   
Although the sample is not flux limited, we find 
a ratio of obscured to unobscured AGN of $1.5 - 2.0$, consistent
with luminosity dependent unified models for AGN. Only four Compton-thick AGN are found in the sample. This implies that the missing Compton-thick AGN needed to explain the cosmic hard X-ray background would have to have lower fluxes than discovered by {\it INTEGRAL} so far.
\end{abstract}

\section{Introduction}

The X-ray sky as seen by satellite observations over the past 40 years, shows a 
substantially different picture than for example the optical band. While the visual 
night sky is dominated by main sequence stars, Galactic binary systems and super nova remnants form the brightest objects X-rays. Common to both regimes 
is the dominance of active galactic nuclei (AGN) toward lower fluxes. 
In the X-ray range itself, one observes a slightly different
population of AGN at soft and at hard X-rays. Below 5 keV the X-ray
sky is dominated by AGN of the Seyfert 1 type; above 5 keV the absorbed
Seyfert 2 objects appear to become more numerous. 
These type 2 AGN are also believed to be the main contributors to the cosmic X-ray
background above 5 keV (Setti \& Woltjer 1989; Comastri et al. 1995; Gilli et al. 2001), although 
only $\sim 50 \%$ of the XRB above $8 \rm \, keV$ can be resolved (Worsley et al. 2005).

The hard X-ray energy range is not currently accessible to X-ray telescopes using grazing 
incidence mirror systems. Instead detectors without spatial resolution like the 
PDS on {\it BeppoSAX} and OSSE on {\it CGRO} 
have been applied. 
A synopsis of these previous results is as follows: the 2 -- 10 keV Seyfert~1 continua 
are approximated by a $\Gamma \simeq 1.9$ powerlaw form (Zdziarski et al. 1995). A flattening 
above $\sim 10 \, \rm keV$ has been noted, and is commonly attributed to Compton reflection (George \& Fabian 1991). 
There is a great deal of additional detail in this spectral domain - ``warm'' absorption, 
multiple-velocity component outflows, and relativistic line broadening - which are beyond the scope of this paper. The Seyfert~2 objects are more poorly categorized here, but the general 
belief is that they are intrinsically equivalent to the Seyfert~1s, but viewed through much 
larger absorption columns.

Above 20 keV the empirical picture is less clear. The $\sim 20 - 200 \, \rm keV$ continuum shape of both Seyfert types is consistent with a thermal Comptonization spectral form, 
although in all but a few cases the data are not sufficiently constraining to rule out 
a pure powerlaw form. Nonetheless, the non-thermal scenarios with pure powerlaw continua extending to $\sim \, \rm MeV$ energies reported in the pre-{\it CGRO} era are no longer 
widely believed, and are likely a result of background systematics. However, a detailed 
picture of the Comptonizing plasma - its spatial, dynamical, and thermo-dynamic 
structure - is not known. Among the critical determinations which {\it INTEGRAL} or future hard 
X-ray instruments will hopefully provide are the plasma temperature and optical depth 
(or Compton ``Y'' parameter) for a large sample of objects. 

The other major class of gamma-ray emitting AGN - the blazars (FSRQs and BL Lac objects) are 
even more poorly constrained in the {\it INTEGRAL} spectral domain (for early {\it INTEGRAL} results see for example
Pian et al. 2005). 

Critical to each of these issues is the need to obtain improved continuum measurements over 
the hard X-ray to soft gamma-ray range for as large a sample of objects as possible. 
{\it INTEGRAL}, since its launch in October 2002, offers unprecedented 
$> 20 \, \rm keV$ collecting area and state of the art detector electronics and background 
rejection capabilities. Thus it offers hope of substantial gains in our knowledge of the AGN phenomenon and in particular of the cosmic hard X-ray background. 

The first {\it INTEGRAL} AGN catalogue offers the possibility to address these questions and to compare the results with previous missions.




\section{The {\it INTEGRAL} AGN Sample}

\begin{figure*}
\centering
\epsfig{file=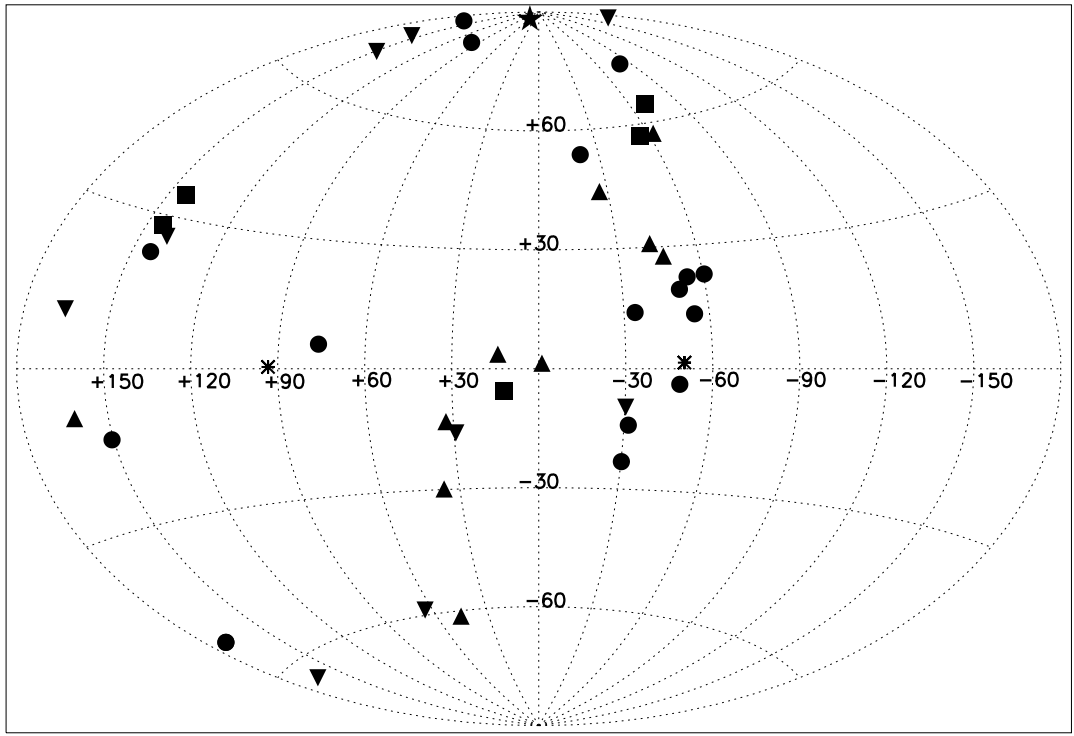,width=0.85\linewidth}
\caption{The distribution of {\it INTEGRAL} AGN in the sky in Galactic coordinates. Seyfert 1 are marked with up-triangles, Seyfert 1.5 with down-triangles, Seyfert 2 with circles, blazars with squares, optically unidentified with asterisks, and the Coma Cluster is represented by a star.\label{fig:agnmap}}
\end{figure*}

Our {\it INTEGRAL} AGN sample\footnote{http://heasarcdev.gsfc.nasa.gov/docs/integral/spi/pages/agn.html} consists of 42 extragalactic objects, detected in the $20 - 40 \rm \, keV$ energy band with the imager IBIS/ISGRI. Spectra have been extracted from IBIS/ISGRI, the spectrometer SPI, and the X-ray monitor JEM-X in order to cover the energy range from $3 - 500 \, \rm keV$. The list of sources with their redshift, the optical counterpart type, the flux in the 20 -- 40 keV band as measured by ISGRI, the luminosity in the 20 -- 100 keV band, and the intrinsic absorption as measured at soft X-rays is given in Table~\ref{tab:sample}. Details on the analysis and on individual spectra can be found in Beckmann et al. (2006). 
The distribution of sources in the sky is shown in Figure~\ref{fig:agnmap}. 

\begin{figure}
\centering
\epsfig{file=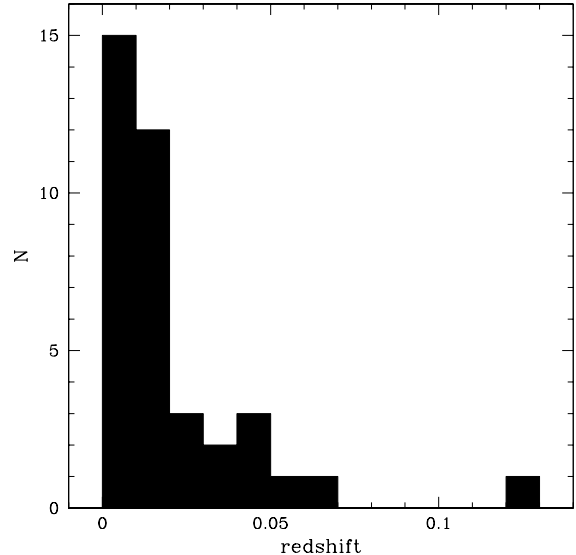,width=1.0\linewidth}
\caption{Redshift distribution of the AGN detected by {\it INTEGRAL}. Blazars are not shown. The average redshift is ${\bar z} = 0.020$. The object on the right is the quasar PG~1416--129.\label{fig:zhisto}}
\end{figure}

The Seyfert type AGN found in the sample are preferentially low redshift objects. Figure~\ref{fig:zhisto} shows the distribution of redshifts in the sample. The blazars all show higher redshifts ($0.15 < z < 2.51$) and are not included in the histogram. The one object on the right is PG~1416--129 ($z = 0.1293$) and is an anomalous radio quiet quasar with similar spectroscopic properties as radio-loud sources (Sulentic et al. 2000). 


In order to investigate the AGN subtypes, we have derived
averaged spectra of the Seyfert 1 and 2 types, as well as for the
intermediate Seyferts and the blazars.
The average Seyfert 1 spectrum was constructed using the weighted mean
of 10 ISGRI spectra, the Seyfert 2 composite spectrum includes 15
sources, and 8 objects form the intermediate Seyfert 1.5 group. The
two brightest sources, Cen~A and NGC~4151, have been excluded from the
analysis as their high signal-to-noise ratio would dominate the
averaged spectra.
The average spectra have been constructed by computing the weighted mean of all fit results on the individual sources. 
In order to do so, all spectra had been fit by an absorbed single
powerlaw model. When computing the weighted average of the various
sub-classes, the Seyfert 2 objects show flatter hard X-ray spectra
($\Gamma = 1.95 \pm 0.01$) than the Seyfert 1.5 ($\Gamma = 2.10 \pm
0.02$), and Seyfert 1 appear to have the steepest spectra ($\Gamma =
2.11 \pm 0.05$) together with the blazars ($\Gamma = 2.07 \pm 0.10$). 

The Seyfert type classification of the objects is
based on optical observations. An approach to classifying sources according to
their properties in the X-rays can be done by separating the sources
with high intrinsic absorption ($N_{\rm H} > 10^{22} \rm \, cm^{-2}$)
from those objects which do not show significant absorption. 
\begin{figure}
\centering
\epsfig{file=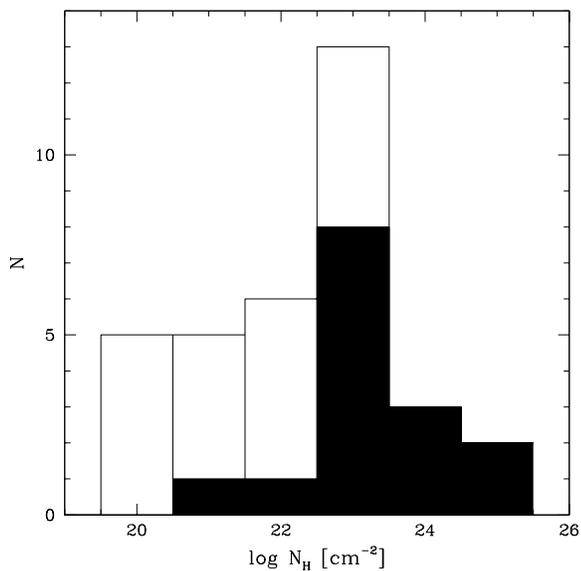,width=1.0\linewidth}
\caption{Distribution of intrinsic absorption, as measured in the soft X-rays. The Seyfert 2 objects (including the Seyfert 1.8 and 1.9 subtypes) are shown in black.\label{fig:nh_histogram}}
\end{figure}
The distribution of absorption at soft X-rays for the AGN sample is shown in Figure~\ref{fig:nh_histogram}. The black part of the histogram represents the Seyfert 2 AGN (including the Seyfert 1.8 and Seyfert 1.9 subtypes). 
It has to be pointed out that not all objects which show
high intrinsic absorption in the X-rays are classified as Seyfert~2
galaxies in the optical, and the same applies for the other AGN
sub-types. Nevertheless a similar trend in the spectral slopes can be
seen: the 21 absorbed AGN show a flatter hard X-ray spectrum ($\Gamma
=1.98 \pm 0.01$) than the 13 unabsorbed sources ($\Gamma
=2.08 \pm 0.02$). 
The blazars have again been excluded from these
samples. 

Among the Seyfert~2 galaxies we find 4 objects with low absorption ($N_H < 10^{23} \rm \, cm^{-2}$), 7 with intermediate absorption  ($N_H = 10^{23} - 10^{24}  \rm \, cm^{-2}$), and four Compton thick AGN ($N_H > 10^{24} \rm \, cm^{-2}$).

\begin{figure}
\centering
\epsfig{file=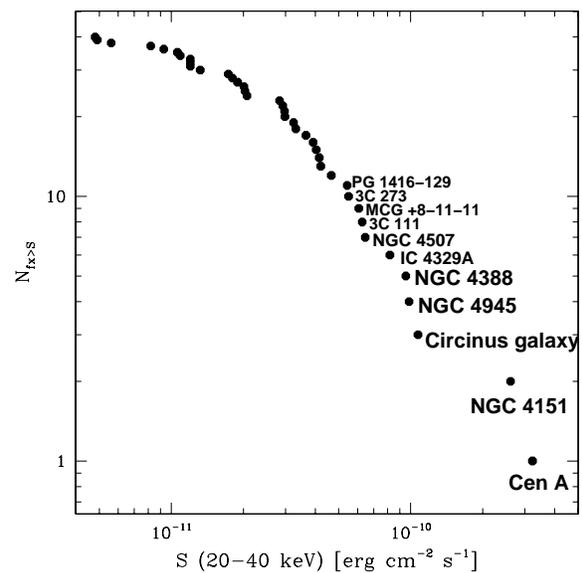,width=1.0\linewidth}
\caption{Number counts of the INTEGRAL AGN sample. The brightest objects have been labelled.\label{fig:logNlogS}}
\end{figure}

Although the {\it INTEGRAL} AGN sample discussed here is not a complete flux limited one, the number counts give a first impression regarding the flux distribution within the sample (Fig.~\ref{fig:logNlogS}). 
Excluding the two brightest objects (Cen~A and NGC~4151) and the objects with $f_X < 2 \times 10^{-11} \rm \, erg \, cm^{-2} \, s^{-1}$ (where the number counts shows a turnover), the number counts relation shows a gradient of $1.4 \pm 0.1$, and is consistent with the value of 1.5 expected for Eucledian geometry and no evolution in the local universe.

\begin{figure}
\centering
\epsfig{file=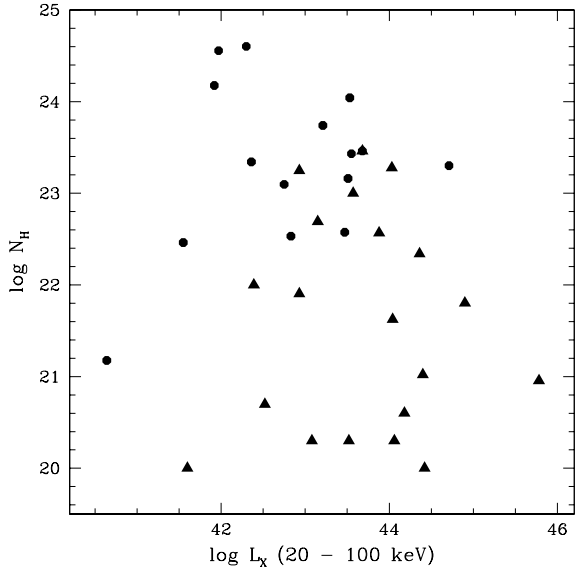,width=1.0\linewidth}
\caption{Intrinsic absorption, as measured at soft X-rays, versus the luminosity in the 20 -- 100 keV band. Circles represent Seyfert 2 type AGN (including Sy~1.8 and Sy~1.9), triangles other Seyfert types.\label{fig:logLx_NH}}
\end{figure}

Figure~\ref{fig:logLx_NH} shows the intrinsic absorption of the objects, as measured in the soft X-rays, versus the luminosity in the 20 -- 100 keV energy band as measured by {\it INTEGRAL}. There is no discernable correlation between those values. The optically classified Seyfert 2 objects naturally populate the area with higher intrinsic absorption.

\section{Discussion}

The typical {\it INTEGRAL} spectrum can be described by a simple
powerlaw model with average photon indices ranging from $\Gamma = 2.0$ for
obscured AGN to $\Gamma = 2.1$ for unabsorbed sources. The simple model does not give reasonable results in high signal-to-noise cases, where an appropriate fit requires additional features such as a cut-off and a reflection component (Soldi et al. 2005, Beckmann et al. 2005).
The results presented here show slightly steeper spectra than previous
investigations of AGN in comparable energy ranges. The same trend is
seen in the comparison of Crab observations, where the {\it
  INTEGRAL}/ISGRI spectra also appear to be slightly steeper than in
previous observations, and in comparison of {\it RXTE} and {\it
  INTEGRAL}/ISGRI spectra of Cen~A (Rothschild et al. 2006). This trend might be based on improper calibration and/or background subtraction.

The average properties, like spectral slope, redshift, luminosity, are rather similar compared to previous studies (e.g. Zdziarski et al. 1995, Gondek et al. 1996). 

Comparing the ratio $X$ of obscured ($N_{\rm H} > 10^{22} \rm \, cm^{-2}$) to
unobscured AGN we find in the {\it INTEGRAL} data that $X = 1.7 \pm
0.4$. 
The ratios change slightly when taking into account only those
objects which belong to the complete sample with an ISGRI significance
of $7 \sigma$ or higher (Beckmann et al. 2006). This sub-sample includes
32~AGN, with 18 obscured and 10 unobscured objects (absorption
information is missing for the remaining four objects).
Using only the complete sample gives a similar ratio of 
$X = 1.8 \pm 0.5$.
Splitting this result up into objects near the Galactic
plane ($|b| < 20^\circ$) and off the plane shows for all objects a
ratio of $X = 3.3 \pm 1.1$ and $X = 1.1 \pm 0.5$, respectively. This
trend shows that the harder spectra of those objects, where the absorption in the line of sight through the Galaxy is low compared to the intrinsic absorption, are more likely to shine through the Galactic plane.

Risaliti et al. (1999) studied a large sample of Seyfert 2 galaxies focusing especially on the intrinsic absorption measured at soft X-rays. They also find a fraction of 75\% of Seyfert~2 with an intrinsic absorption $N_H > 10^{23} \rm \, cm^{-2}$, but a  50\% fraction of Compton-thick objects with $N_H > 10^{24} \rm \, cm^{-2}$, where the {\it INTEGRAL} sample only finds 4 objects (27 \%).

Optical studies in the local universe find evidence that type 2 AGN are about a factor of four more numerous than type 1 AGN (Setti \& Woltjer 1989; Comastri et al. 1995). In X-rays the situation is similar, although not all Seyfert 1 objects show low intrinsic absorption and vice versa (see Fig.~\ref{fig:logLx_NH}). Recent studies have shown that the fraction of absorbed sources depends both on luminosity and redshift in a way that the fraction of type 2 AGN increases towards higher redshifts and lower luminosity (e.g. Gilli et al. 2001; Ueda et al. 2003; La Franca et al. 2005). Ueda et al. (2003) sudied 247 AGNs in the 2 -- 10 keV band with luminosities in the range $L_X = 10^{41.5} - 10^{46.5} \rm \, erg \, s^{-1}$, similar to the luminosity range of the {\it INTEGRAL} AGN but extending to higher redshifts ($z = 0 - 3$) and to fainter fluxes ($f_X = 10^{-10} - 3.8 \times 10^{-15} \, \rm erg \, cm^{-2} \, s^{-1}$). They find that the number of AGN decreases with the intrinsic X-ray luminosity of the AGN and therefore favour a luminosity dependend density evolution (LDDE) to explain the luminosity function in the $2 - 10 \rm \, keV$ band. 
La Franca et al. (2005) used an even larger sample and confirmed the necessity of a LDDE model, where low luminosity AGN peak at $z \sim 0.7$, while high luminosity AGN peak at $z \sim 2.0$. In addition, they find evidence that the fraction of absorbed ($N_H > 10^{22} \rm \, cm^{-2}$) AGN decreases with the intrinsic luminosity in the $2 - 10 \rm \, keV$ energy range. Consistent with our study, La Franca et al. also find a ratio of $X = 2.1$ at $L_X = 10^{42.5} \rm \, erg \, s^{-1}$.

All these results based on the 2 -- 10 keV band are consistent with the findings of the {\it INTEGRAL} AGN sample at higher energies ($20 - 40 \rm \, keV$). It is surprising though that different from the findings of Risaliti et al. (1999) we do not detect a large fraction of Compton-thick AGN. These AGN, if exisiting, could explain the peak in the hard X-ray background around 30 keV (e.g. Maiolino et al. 2003). In view of recent results this lack of Compton-thick objects is explainable by the type of AGN detected so far by {\it INTEGRAL}. All objects are local AGN, with a mean redshift of ${\bar z} = 0.020$ (Fig.~\ref{fig:zhisto}). The objects are bright ($f_{20 - 40 \rm \, keV} > 5 \times 10^{-12}$; Fig.~\ref{fig:logNlogS}), but have low luminosities (${\bar L_X} = 2 \times 10^{43} \rm \, erg \, s^{-1}$). 
This still leaves room in the parameter space of AGN to locate Compton-thick AGN. Those objects could have lower fluxes and therefore even lower luminosities in the local universe than the objects studied by {\it INTEGRAL}. This possibility is supported by the unified model for AGN as described by Treister \& Urry (2005). They predict a strong correlation of the fraction of broad line AGN with luminosity, and expect up to a factor of 10 more absorbed than un-absorbed AGN at very low luminosities ($L_X \simeq 10^{42} \rm \, erg \, s^{-1}$).
This trend can be seen also in the {\it INTEGRAL} AGN sample. When considering only the 14 objects with $L_{20 - 100 \rm \, keV} < 10^{43} \rm \, erg \, s^{-1}$ the ratio of obscured to unobscured objects increases to $X = 2.5$. 

Studying the population of sources at lower limiting fluxes should also reveal further highly absorbed sources at high redshifts, because these have been missed so far and with increasing redshift an increase of absorbed AGN fraction is expected (La Franca et al. 2005).

The energy range in the 15 -- 200 keV is now also accessible through
the BAT instrument aboard {\it Swift} (Gehrels et al. 2005). A study by
Markwardt et al. (2005) used data from the first three months of the {\it Swift} mission for studying the extragalactic sky and reached a flux limit of $f_{(14 - 195 \rm \, keV)} \simeq 10^{-11} \rm \, erg \, cm^{-2} \, s^{-1}$. The source population is similar to the {\it INTEGRAL} one, with an average redshift of ${\bar z} = 0.012$, and a ratio of $X = 2$
between obscured and unobscured AGN, fully consistent with {\it INTEGRAL}. Although the {\it Swift}/BAT survey
covers different areas of the sky, the energy band is similar to the
{\it INTEGRAL}/ISGRI range and the type of AGN detectable should be
the same. Within their sample of 44 AGN they detect 5 Compton-thick AGN, the same ratio as in the {\it INTEGRAL} sample. 
The only difference appears in the relation between luminosity and absorption. While in the {\it INTEGRAL} sample no correlation is detectable (Fig.~\ref{fig:logLx_NH}), Markwardt et al. find in their sample evidence for an anti-correlation of luminosity and absorption.


\section{Conclusions}

The {\it INTEGRAL} AGN sample opens the window to the hard X-ray sky above 20 keV for population studies. With the 42 extragalactic objects discussed here, a fraction of about $60\%$ shows absorption above $N_H = 10^{22} \rm \, cm^{-2}$, but only four objects are actually Compton-thick ($N_H > 10^{24} \rm \, cm^{-2}$). This shows that the source population above 20 keV, at least at the high flux end, is very similar to the one observed in the $2 - 10 \rm \, keV$ energy region. The results are consistent with observations by the {\it Swift}/BAT instrument, although we cannot confirm a correlation of X-ray luminosity with instrinsic absorption. 
Further investigations are necessary and with the ongoing {\it INTEGRAL} and {\it Swift} mission it will be revealed in the near future, if there is a significant Compton-thick AGN population to explain the peak in the extragalactic X-ray background around 30 keV. These objects would have to have lower fluxes than the objects studied here (i.e. $f_X < 10^{-11} \rm \, erg \, cm^{-2} \, s^{-1}$) and might therefore be low-redshift, low-luminosity ($L_X \lae 10^{42} \rm \, erg \, s^{-1}$) Seyfert galaxies, or higher redshift ($z \gg 0.05$) objects.

\begin{table}
  \begin{center}
    \caption{The {\it INTEGRAL} AGN sample}\vspace{1em}
    \renewcommand{\arraystretch}{1.2}
    \begin{tabular}[h]{lllccc}
      \hline
      Name & z & type & $f_{(20 - 40 \rm \, keV)}$ & $\log L_{(20 - 100 \rm \, keV)}$ & $N_H$\\
	&   &      & $[10^{-11} \rm \, erg \, cm^{-2} \, s^{-1}]$ & $[\rm erg \, s^{-1}]$ & $[10^{22} \rm \, cm^{-2}]$ \\
      \hline
NGC 788  & 0.0136 & Sy 1/2 & $2.98 \pm 0.24$ & 43.52 & $< 0.02$ \\
NGC 1068 & 0.0038 & Sy 2 & $0.93 \pm 0.27$ & 41.92 & $> 150$\\
NGC 1275 & 0.0176 & Sy 2 & $1.89 \pm 0.21$ & 43.47 & 3.75\\
3C 111   & 0.0485 & Sy 1 & $6.27 \pm 0.57$ & 44.90 & 0.634\\
MCG +8--11--11 & 0.0205 & Sy 1.5 & $6.07 \pm 0.97$ & 44.06 & $< 0.02$\\
MRK 3    & 0.0135 & Sy 2 & $3.65 \pm 0.39$ & 43.53 & 110 \\ 
MRK 6    & 0.0188 & Sy 1.5 & $2.01 \pm 0.20$ & 43.57 & 10 \\
NGC 4051 & 0.0023 & Sy 1.5 &$1.80 \pm 0.20$ & 41.60 & $<0.01$ \\
NGC 4151 & 0.0033 & Sy 1.5 &$26.13 \pm 0.16$& 43.15 & 4.9 \\
NGC 4253 & 0.0129 & Sy 1.5 &$0.93 \pm 0.22$ & 42.93 & 0.8\\
NGC 4388 & 0.0084 & Sy 2 & $ 9.54 \pm 0.25$ &43.55 & 27\\
NGC 4395 & 0.0011 & Sy 1.8 &$0.56 \pm 0.22$ & 40.64 & 0.15\\
NGC 4507 & 0.0118 & Sy 2 &$6.46 \pm 0.36$ & 43.68 & 29\\
NGC 4593 & 0.0090 & Sy 1 & $3.31 \pm 0.16$ & 43.08 & 0.02 \\
Coma Cluster & 0.0231 &GClstr. & $1.09 \pm 0.12$ & 43.40 & $<0.01$ \\
NGC 4945 & 0.0019 & Sy 2 &$ 9.85 \pm 0.23$ &42.30 & 400\\
ESO 323--G077 & 0.0150 & Sy 2 &$1.20 \pm 0.19$ &43.21 & 55\\
NGC 5033 & 0.0029 & Sy 1.9 & $1.06 \pm 0.24$ & 41.55 & 2.9\\
Cen A    & 0.0018 & Sy 2 &$32.28 \pm 0.17$ &42.75 & 12.5\\
MCG--06--30--015 & 0.0077 & Sy 1 & $2.98 \pm 0.19$ & 42.93 &17.7 \\
4U 1344--60 & 0.043 & ? & $2.83 \pm 0.15$ &44.36 &2.19 \\
IC 4329A & 0.0161 & Sy 1.2 & $8.19 \pm 0.17$ & 44.04 & 0.42 \\
Circinus gal.& 0.0014 & Sy 2 &$10.73 \pm 0.18$ & 41.97 & 360 \\
NGC 5506 & 0.0062 & Sy 1.9 & $4.21 \pm 0.33$ & 42.83 & 3.4\\ 
PG 1416--129 & 0.1293 & Sy 1 & $5.43 \pm 0.64$ & 45.78 &0.09 \\
IC 4518 & 0.0157 & Sy 2 &$0.49 \pm 0.32$ & 42.92 & ? \\
NGC 6221 & 0.0050 & Sy 1/2 & $1.32 \pm 0.20$ & 42.39 & 1\\
NGC 6300 & 0.0037 & Sy 2 &$3.91 \pm 0.37$ & 42.36 & 22\\
GRS 1734--292 & 0.0214 & Sy 1 & $4.03 \pm 0.09$ & 43.88 & 3.7\\
IGR J18027--1455& 0.0350 & Sy 1 & $2.03 \pm 0.16$ & 44.03 & 19.0\\
ESO 103--G35 & 0.0133 & Sy 2 &$2.97 \pm 0.66$ & 43.51 & 13 -- 16\\
1H 1934--063& 0.0106 & Sy 1 & $0.48 \pm 0.25$ & 42.51 & ? \\
NGC 6814 & 0.0052 & Sy 1.5 & $2.92 \pm 0.23$ & 42.52 & $< 0.05$\\
Cygnus A & 0.0561 & Sy 2 &$3.24 \pm 0.14$ & 44.71 & 20\\
MRK 509  & 0.0344 & Sy 1 & $4.66 \pm 0.47$ & 44.42 & $< 0.01$ \\
IGR J21247+5058& 0.020 & radio gal.? & $4.15 \pm 0.27$ & 44.00 & ? \\ 
MR 2251--178 & 0.0634 & Sy 1 & $1.20 \pm 0.17$ & 44.40 & 0.02 -- 0.19\\ 
MCG --02--58--022 & 0.0469 & Sy 1.5 & $1.20 \pm 0.28$ &44.18 & $<0.01 - 0.08$\\[0.5cm]

S5 0716+714 & 0.3$^a$ & BL Lac & $0.14 \pm 0.11$ & 45.21$^a$ & $<0.01$\\ 
S5 0836+710 & 2.172 & FSRQ & $1.73 \pm 0.28$ & 47.87 & $0.11$\\ 
3C 273   & 0.1583 & Blazar & $5.50 \pm 0.15$ & 45.92 & $0.5$\\
3C 279   & 0.5362 & Blazar & $0.82 \pm 0.24$ & 46.37 & $0.02 - 0.13$\\
PKS 1830--211& 2.507 & Blazar & $2.07 \pm 0.14$ & 48.09 & $<0.01
- 0.7$\\

      \hline \\
$^a$ tentative redshift\\
      \end{tabular}
    \label{tab:sample}
  \end{center}
\end{table}


\section*{References}

Beckmann V., Shrader C. R., Gehrels N., et al. 2005, ApJ in press, astro-ph/0508327

Beckmann V., Gehrels N., Shrader C. R., \& Soldi S. 2006, ApJ accepted, astro-ph/0510530

Comastri A., Setti G., Zamorani G., Hasinger G. 1995, A\&A, 296, 1


Gehrels N., Chincarini G., Giommi P., et al. 2004, ApJ, 611, 1005

George I. M., \& Fabian A. C. 1991, MNRAS, 249, 352

Gilli R., Salvati M., Hasinger G., 2001, A\&A, 366, 407

Gondek D., Zdziarski A. A., Johnson W. N., et al. 1996, MNRAS, 282, 646

La Franca F., Fiore F., Comastri A., et al. 2005, ApJ in press, astro-ph/0509081

Maiolino R., Comastri A., Gilli R., et al. 2003, MNRAS, 344, L59

Markwardt C. B., Tueller J., Skinner G. K., et al. 2005, ApJL in press, astro-ph/0509860

Pian E., Foschini L., Beckmann V., et al. 2005, A\&A, 429, 427 

Rothschild R. E., Wilms J., Tomsick J., et al. 2006, ApJ submitted 

Setti G., \& Woltjer L. 1989, A\&A, 224, L21

Soldi S., Beckmann V., Bassani L., et al. 2005, A\&A accepted, astro-ph/0509123

Sulentic J. W., Marziani P., Zwitter T., et al. 2000, ApJ, 545, L15

Treister E., Urry C. M., 2005, ApJ in press, astro-ph/0505300

Ueda Y., Akiyama M., Ohta K., Miyaji T. 2003, ApJ, 598, 886

Worsley M. A., Fabian A. C., Bauer F. E., et al. 2005, MNRAS, 357, 1281

Zdziarski A. A., Johnson W. N., Done C., et al. 1995, ApJ, 438, L63


\end{document}